# The meteorite flux of the last 2 Myr recorded in the Atacama desert

A. Drouard[1,2], J. Gattacceca[2], A. Hutzler[3], P. Rochette[2], R. Braucher[2], D. Bourlès[2], ASTER Team[2,*], M. Gounelle[4], A. Morbidelli[5], V. Debaille[6], M. Van Ginneken[6,7,8], M. Valenzuela[9,10], Y. Quesnel[2], R. Martinez[11]

[1]*Aix-Marseille Université, CNRS, LAM, Marseille, France*

[2]*Aix-Marseille Univ., CNRS, IRD, Coll France, INRA, CEREGE, Aix-en-Provence, France*

[3]*LPI, Houston, USA*

[4]*IMPMC, Muséum National d'Histoire Naturelle, Sorbonne Universités, CNRS, UPMC & IRD, 57 rue Cuvier, 75005, Paris, France*

[5]*Laboratoire Lagrange, UMR 7293, Université de la Côte d'Azur, CNRS, OCA, Nice, France*

[6]*Laboratoire G-Time, Université Libre de Bruxelles, Belgium*

[7]*AMGC, Vrije Universiteit Brussel, Belgium*

[8]*Geological Survey of Belgium, Royal Belgian Institute of Natural Science, Belgium*

[9]*SERNAGEOMIN, Santiago, Chile*

[10]*Millennium Institute of Astrophysics MAS, Santiago, Chile*

[11]*Museo del Meteorito, San Pedro de Atacama, Chile*

*Georges Aumaître, Karim Keddadouche*

**ABSTRACT**

The evolution of the meteorite flux to the Earth can be studied by determining the terrestrial ages of meteorite collected in hot deserts. We have measured the terrestrial ages of 54 stony meteorites from the El Médano area, in the Atacama Desert, using the cosmogenic nuclide

$^{36}$Cl. With an average age of 710 ka, this collection is the oldest collection of non fossil meteorites at the Earth's surface. This allows both determining the average meteorite flux intensity over the last 2 Myr (222 meteorites larger than 10 g per km$^2$ per Myr) and discussing its possible compositional variability over the Quaternary period. A change in the flux composition, with more abundant H chondrites, occurred between 0.5 and 1 Ma, possibly due to the direct delivery to Earth of a meteoroid swarm from the asteroid belt.

**INTRODUCTION**

The delivery of extraterrestrial matter to the Earth is controlled by the complex dynamical evolution of the Solar System bodies. The past flux of extraterrestrial matter to the Earth's surface has been studied at different spatial and time scales. Impact craters allow quantifying the long term flux but only for large impactors (e.g., Mazrouei et al. 2019). At the other end of the size spectrum, micrometeorites allow studying the extraterrestrial dust reaching the Earth (e.g., Genge 2008, Heck et al. 2017). Meteorites allow estimating the flux of intermediate size (cm-m scale) meteoroids to the Earth's surface. Fossil meteorites may light on the very ancient flux (e.g., Schmitz et al. 2001, Schmitz 2013 for a 1.75 Myr time window during the Middle Ordovician). Meteor observations have allowed estimating the intensity of the current meteorites flux to the Earth (e.g., Halliday et al. 1989, Zolensky et al. 2006), while observed meteorite falls (1312 registered meteorites as of July 2018) provide information about its composition over the last two centuries. On the other hand, meteorite "finds" (whose fall has not been observed, 66165 meteorites as of July 2018) allow constraining the intensity and composition of the meteorite flux on longer time scales more relevant to geological and astronomical processes (e.g. Bland et al. 1996, Benoit et al. 1996, Graf et al. 2001). Most meteorite finds come from Antarctica (64%) and hot deserts (~30%), that are suitable areas for both preservation and recovery. The terrestrial ages of hot desert meteorites are usually in the 0-30 ka range and rarely exceed 50 ka (Jull 2006). Antarctic meteorites have higher terrestrial ages, variable between icefields, but they rarely exceed 150 ka (Welten et al. 2006). These timescales are still short with respect to those involved in dynamical evolution of the Solar System bodies (e.g. Bottke et al. 2005). Moreover, the large Antarctic meteorite collection, that have older terrestrial ages, cannot be easily used to constrain the meteorite flux due to biases introduced by meteorite concentration mechanisms (Williams et al. 1983), and the difficulty in identifying paired fragments (Zolensky et al. 1998), as opposed to passive in situ accumulation in hot deserts. The Atacama Desert is the oldest and driest of hot deserts (Clarke 2005, Dunai et al. 2005). It has been shown to be an important meteorite reservoir, with the highest meteorite density ever determined in hot deserts (Gattacceca et al. 2011, Hutzler et al. 2016). We present in this study the terrestrial ages of 54 meteorites recovered from the Atacama Desert.

**SAMPLES**

We selected randomly 54 unpaired meteorites within the 388 that were found in the El Médano and Caleta el Cobre dense collection areas (Figure 1). These two adjacent areas are collectively called El Médano area in the following. These areas have been shown to bear the highest meteorite concentration in hot deserts (Hutzler et al. 2016). These 54 meteorites are all ordinary chondrites. Ordinary chondrites are the most abundant class of meteorites, and are overrepresented in the El Médano collection (where they represent 96% of the meteorite population) because of recovery biases (Hutzler et al. 2016). The 54 meteorites span the three groups of ordinary chondrites: H (25 meteorites), L (26 meteorites), and LL (3 meteorites).

Based on recovery location, petrography, silicate geochemistry and magnetic susceptibility, special care was taken to discard possibly paired meteorites to avoid statistical overrepresentation of large falls that can produce multiple meteorites. Terrestrial ages were determined by measuring the $^{36}$Cl (half-life: 301±0.01 kyr) concentration in the FeNi metal fraction of these meteorites (see appendix: methods and the supplementary material for the complete data).

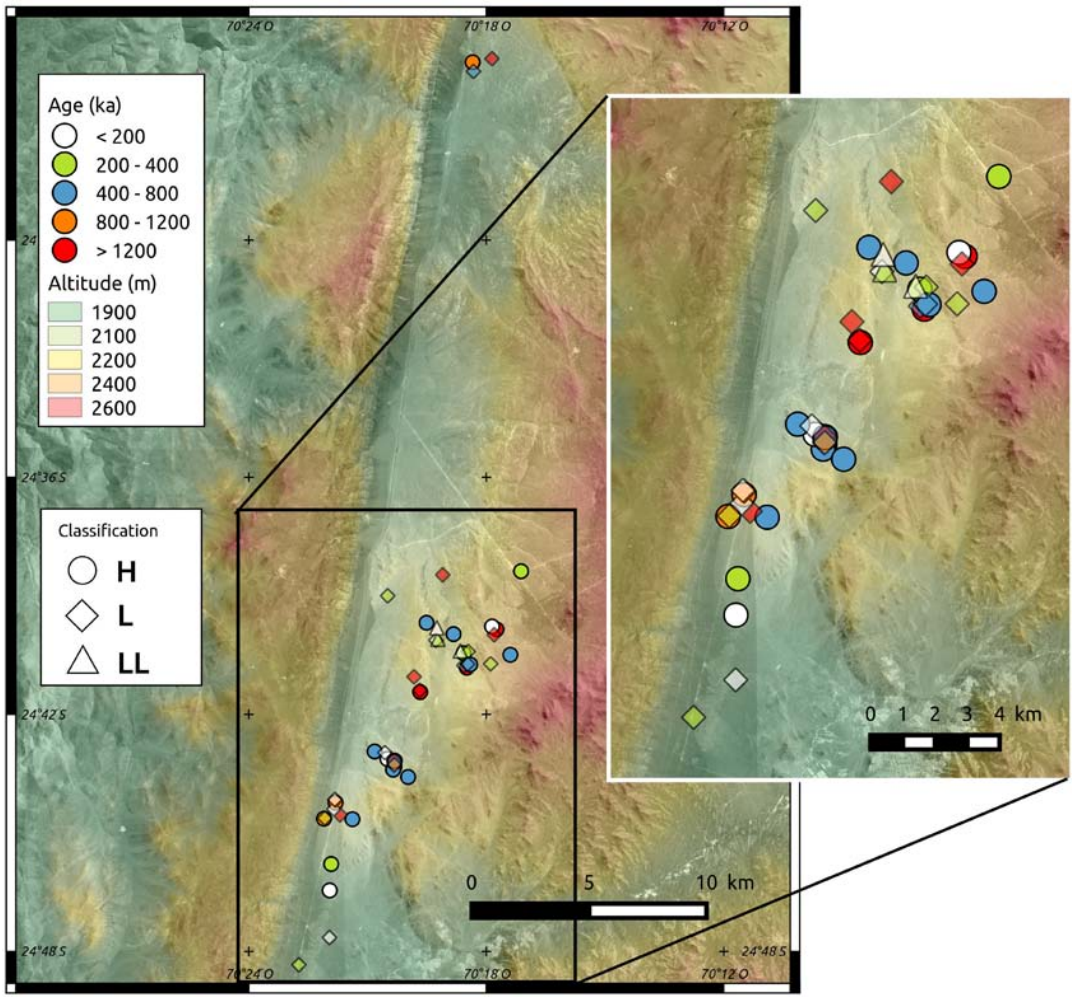

Figure 1. Map of the El Médano and Caleta el Cobre areas with the 54 dated samples, with their class (H, L, or LL) and their age.

**THE OLDEST COLLECTION OF NON-FOSSIL METEORITES**

The absence of geographical trend in the ages (Figure 1) confirms that pairing does not affect the age distribution, and was properly assessed by Hutzler et al. (2016). As an extra precaution, the few groups of meteorites with similar classification and terrestrial age were checked again for possible pairing using a variety of criteria, in particular petrography. No pairing was evidenced. Although shielding in large meteoroids may account for overestimation of some ages, measured cosmogenic radionuclides content of ordinary chondrite falls show that shielding has affected only 3 out of 31 studied meteorites (Graf et al. 2001, Dalcher et al. 2013). Furthermore, the meteorite fall population is strongly biased towards large masses (more likely to be observed and recovered), as evidenced by their mass distribution compared to that of meteorite finds or theoretical estimates (e.g., Huss 1990). We note that the 3 ordinary chondrites showing a significant shielding (Richardton, Uberaba and La Criolla meteorites) all have masses above 40 kg. Such large meteorites are exceedingly rare in unbiased meteorites collection like the El Médano collection. Therefore, only a few percent of the El Médano meteorite collection may have been affected by shielding. This would correspond to a couple of meteorites with no consequence on the overall age distribution. With a mean terrestrial age of 710 ka, the El Médano collection is by far the oldest collection of non-fossil meteorites on Earth's surface (Figure 2). About 30% of the samples are older than 1 Ma, and two are older than 2 Ma.

For comparison, meteorites from other hot deserts and Antarctica have an average terrestrial age of only 12 ka and 99 ka respectively (computed on 152 and 398 terrestrial ages from MetBase database (Koblitz 2005), in agreement with Jull (2006)). The results of the present study are consistent with the high ages of the Atacama desert surfaces associated to long-standing hyperaridy (Clarke 2005, Dunai et al. 2005), and offer an explanation for the unusually high number of meteorites that were found in the El Médano area (~190 unpaired meteorites larger than 10 g per km² (Hutzler et al. 2016)). The age distribution can be fitted with an exponential law decrease with a half-life 590 kyr (Figure 2). The smooth exponential decrease suggests that no discrete meteorite removal event, such as surface reworking, occurred over the last 2 Myr, but instead, that meteorites are removed by continuous processes, likely wind abrasion and fragmentation. This distribution suggests that ordinary chondrites are unlikely to be preserved for more than 2.5-3 Myr in the Atacama Desert, except perhaps for the large stones (>10 kg) that are exceedingly rare.

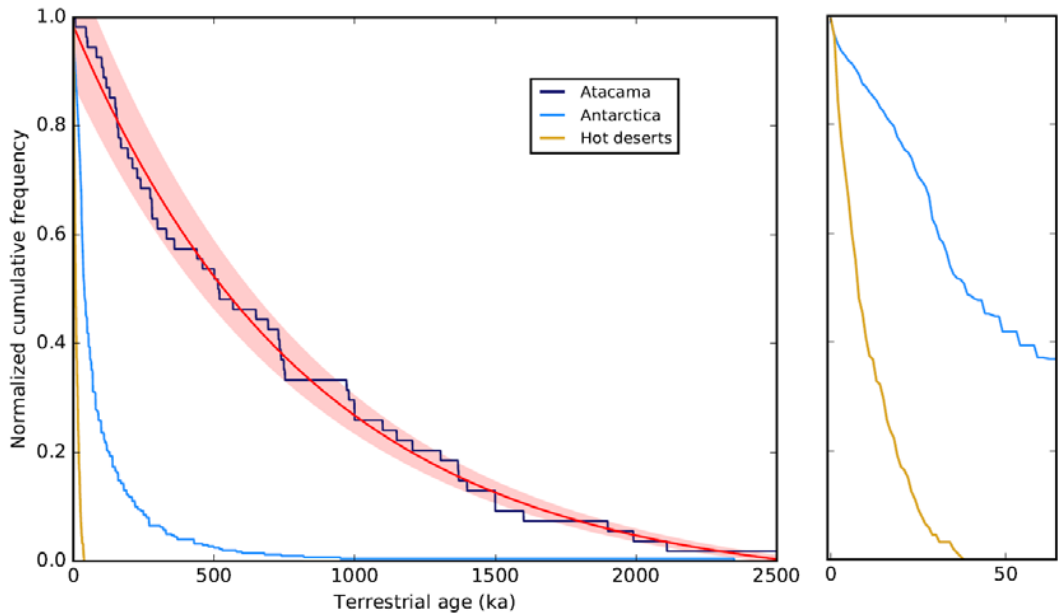

Figure 2. Cumulative terrestrial age distribution measured by $^{36}$Cl (blue) and the exponential best-fit (red). The right panel is a focus on the range 0-65 ka. The shaded area corresponds to the maximum age uncertainty of 90 ka, derived by propagation. Our results are compared to the cumulative age distribution of Antarctic (blue) and hot desert (yellow). Both $^{14}$C and $^{36}$Cl age measurements were selected for Antarctica meteorites, whereas the data for hot desert meteorites (Atacama samples excluded) only consist of $^{14}$C measurements.

# METEORITE FLUX ESTIMATE

The smooth decrease of the age distribution is consistent with a constant meteorite flux combined with meteorite removal by weathering. This assumption means that the resulting meteorite surface density N(t) on the ground satisfies the following evolution equation:

$$dN(t) = \alpha * dt - \lambda * N(t) * dN(t)$$

where $\alpha$ is the meteorite flux and $\lambda$ is the decay rate ($\lambda = \ln(2)/t_{1/2}$). This equation can be solved:

$$N(t) = (1 - e^{-\lambda t}) * \alpha/\lambda + N(t=0)$$

Whatever the value of N(t=0), an equilibrium state is reached after a few half-lives (590 kyr), after which the meteorite surface density remains constant at a saturated value $N_{sat} = \alpha/\lambda$. We consider here meteorites with mass higher than 10g. Using the meteorite density of 189±14 km$^{-2}$ according to Hutzler et al. (2016) and the decay rate computed analytically from the terrestrial age distribution (Figure 2), we determined a fall rate of 222±15 meteorites above 10 g per km$^2$ per Myr. This estimate is similar to the 225±81 meteorites above 10 g per km$^2$ per Myr over the last 50 kyr, given in Bland et al. (1996) by dating meteorites from Nullarbor (Australia). However, these values are higher than the estimate of 83 meteorites above 10 g per km$^2$ per Myr determined by Halliday et al. (1989). This latter estimate is less representative of the average meteorite flux because of the low number of events (56 meteorite dropping events versus 388 meteorites in this work) and the short time window (11 years versus 2 Myr in this work).

# METEORITE FLUX VARIABILITY DURING THE QUATERNARY?

H and L chondrites represent the vast majority (78%) of meteorites and offer the most robust statistical indicator of possible variation in the composition of the meteorite flux. The H chondrite fraction with respect to the total (H+L) chondrites is presented in Figure 3. The overall

H/L ratio of our sample selection (25/26) is different from the H/L ratio of the total El Médano collection (1.74 after pairing, Hutzler et al. 2016). To correct for this bias introduced by our random selection, we applied scaling factors to the H and L numbers so that the overall H/L ratio is 1.74 while the total number of H+L meteorites remains 51. The error bars were computed as the standard deviation of a Poisson distribution, that describes the number of independent events occuring in a interval of time. The plot shows that H chondrites dominated the flux between 0.4 and 1.2 Ma. A change, leading to nearly equal proportion of H and L, occurs at 0.5 Ma and broadly fits with the falls and other hot deserts finds proportions (Figure 3). Previous studies have shown that H chondrites weather more rapidly than L chondrites most likely because of their higher metallic iron content that is more sensitive to weathering (Bland et al.1998 ; Munayco et al.2013). This effect may account for the higher proportion of L chondrites before 1 Ma but contradicts the observed higher proportion of H chondrite between 0.5 and 1 Ma. Therefore, a dynamical explanation must be invoked. The delivery of meteorites to the Earth from the main-belt starts when the meteoroid orbits are dynamically excited above the Earth-crossing eccentricity threshold by a resonance in the asteroid belt (e.g., Bottke et al. 2000). A population of meteoroids on resonant orbits has a typical lifetime of 0.5 Myr before decaying in number mostly by colliding with the Sun (Gladman et al. 1997). The duration is fully consistent with the timescale of the bump in Figure 3. Consequently, a potential scenario could be the direct delivery of a population of meteoroids in a resonance of the main-belt following the break-up of or cratering event on a close-by asteroid. A population of debris (the meteoroids) entered the resonance about 1 Myr ago, and the eccentricities of the meteoroids increased rapidly so that their trajectories started to intersect the Earth's orbit. The flux of meteorites from this population started to decay after 0.5 Myr. Another hypothesis is that a swarm of meteoroids on very similar orbits, similar to a cometary trail but generated by the break-up of a Near-Earth asteroid (e.g., Wiegert and Brown 2005), had a favorable encounter configuration with the Earth between 0.5 and 1Myr ago. The geometry of intersection of an inclined orbit with the Earth depends on the value of the argument of perihelion relative to the ecliptic. In that way, the precession rate of the argument of perihelion sets the duration of the favorable encounter configuration and the timescale on which the configuration repeats. The typical precession rates of the argument of perihelion of Near Earth Asteroids are in the range 10-90 arcsec per year. This implies that each orbital intersection is short-lived and repeats every 40 kyr. This period is 20 times smaller than the duration of the H-peak evidence by our results, which does not favor this scenario. Both scenarios remain hypothetic and need to be investigated in details, especially by measuring the cosmic-ray exposure age of the dataset. Indeed, a short exposure age for the excess of H chondrites of the bump is required for such a rapid delivery from the asteroid belt to the Earth.

**CONCLUSION**

With a mean terrestrial age of 710 ka, the El Médano meteorite collection is by far the oldest meteorite collection on Earth's surface. This confirms the long-term, multi-Myr stability of the Atacama desert surfaces and offers a unique opportunity to study the meteorite flux to the Earth and meteorite weathering over the Myr time scale. Our results suggest that although the bulk flux has remained constant with about 220 meteorites above 10 g per $km^2$ per Myr, significant changes in the composition of the flux may have occurred over the last 2 Myr. In particular an increase in the proportion of H chondrites is suggested between 0.5 and 1 Ma. Our favored scenario would be the direct delivery of a population of meteoroids from the asteroid main-belt. Measurement of the cosmic ray exposure ages of these meteorites could confirm this hypothesis.

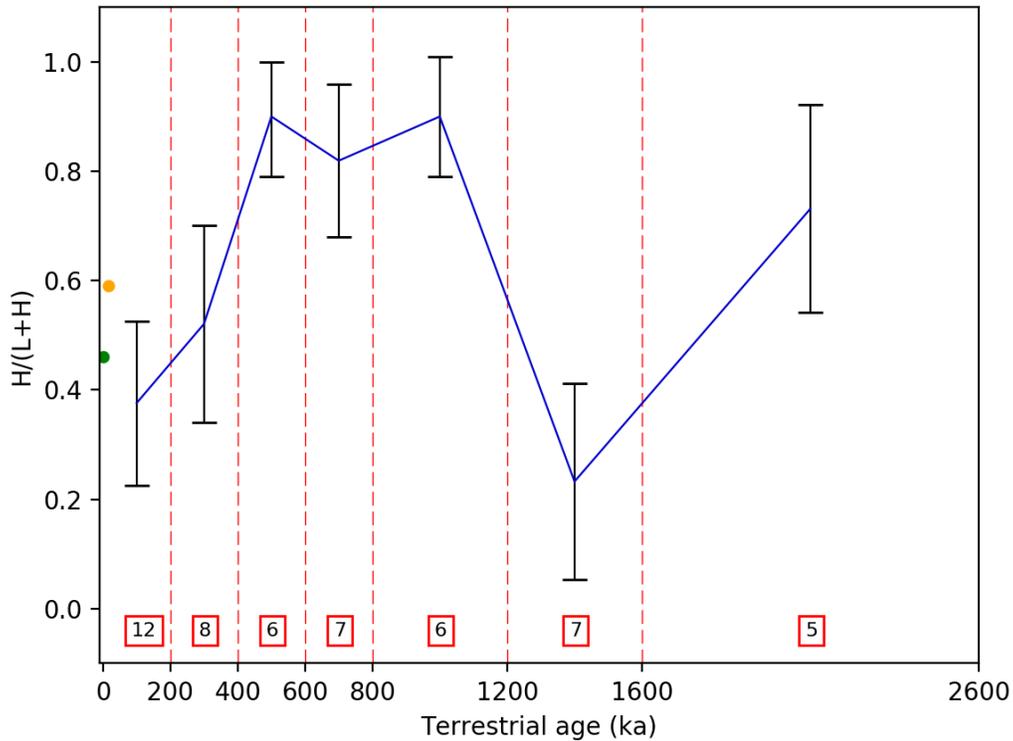

Figure 3. H chondrite fraction (with respect to the total H+L chondrite) evolution during the last 2.6 Myr. The values derived from our dataset have been corrected by scaling factor so that the overall H/L ratio matches the H/L ratio of 1.74 estimated for the whole El Médano collection (see text). Vertical dashed line show the bins, and the numbers in boxes are the number of samples within the bin. Uncertainties were computed as the standard deviation of a Poisson distribution. The green dot is the current ratio computed from the meteorite fall population, whereas the orange dot is the value computed from hot desert collection. Uncertainties are too small to be shown.

**APPENDIX: METHODS**

The complete dating protocol, detailed below, was inspired by the works of Vogt and Herpers 1988 and Merchel and Herpers 1999.

**Metal extraction**

Samples were grinded manually in an agate mortar, and the resulting powder (Ø <200μm) was completed with ethanol and for a two minutes bath in an ultrasonic cuve to break the silicate grains. Once dried, a magnetic separation was performed with a hand magnet. The magnetic extract was then washed in hydrochloric (0.2 mol/L) and hydrofluoric (3.8 mol/L) acids, during 30 and 10 min respectively. The remaining grains were washed, dried, and a final visual inspection with a binocular microscope ensured the purity of the metallic extract.

**Chlorine chemistry**

The metal was then dissolved by nitric acid (2 mol/L) and spiked with an enriched $^{35}Cl$ solution (6.92 mg/g). Silver chloride precipitate was formed by adding 2 drops of silver nitrate (0.57 mol.L$^{-1}$) . We stored the tubes in the darkness to prevent photosensibility effects. The solid phase was centrifuged and washed with milliQ water. To remove all trace of sulfure, we added a saturated solution of $Ba(NO_3)_2$ (0.5 mL) with milliQ water (1mL) and $NH_3$ (2 mL) to form a mixed precipitate of $BaSO_4$ and $BaCO_3$. The supernatant was filtered and completed with $HNO_3$ to form a new AgCl precipitate. We rinsed it with concentrated $HNO_3$ (4 mL) and twice with MilliQ water (5 mL) before the final drying.

**Terrestrial age measurements**

AgCl precipitates were transferred to Nickel cathodes and isotopic ratio $^{36}Cl/^{35}Cl$ measurements were performed at ASTER AMS facility. We derived the $^{36}Cl$ activity $A_{mes}$ for each samples, what is directly related to the terrestrial age:

$$T_{age} = 1/\lambda \ln( A_{sat} / A_{mes} )$$

where $\lambda$ is the $^{36}Cl$ half-life ($\lambda = (301\pm0.01)$ kyr, Bartholomew et al. 1955) and $A_{sat}$ the saturation activity in chondritic meteorites exposed to cosmic rays. This value depends on the elemental composition of the dissolved metal. We therefore measured the bulk composition of all metallic fractions (Fe, Ni) by using the Thermoscientific ICap ICP-OES facility at Laboratoire G-Time at ULB. Some samples samples were analyzed by ICP-MS and ICP-OES at the Service d'Analyse des Roches et Minéraux (SARM, Nancy, France). The purity of the dissolved metal fraction was controlled by measuring the mass of dissolved Si for 23 samples: it is only 1‰ of the Fe+Ni mass on average, and 3‰ maximum. These compositions were then used as inputs in the model of Leya (2009) to compute the saturated activities assuming exposure in space for longer than 3 Myr, and meteoroid radii < 20 cm (these two conditions ensure that the $^{36}Cl$ content of the meteorite had reached saturation and that there was no significant shielding). The first assumption is validated by the compilation of chondrite exposure ages (e.g. Graf and Marti 1995), typically more than 4 Ma (> 10 $\lambda$). The second assumption is validated by the size distribution of meteoroids hitting the Earth: using a consensual mass distribution for meteorites hitting the Earth's surface (Huss 1990), and assuming a mass loss of 99% during atmospheric entry, meteoroids with radius > 20 cm (about 110 kg) represent less than 2% of meteoroids with terminal masses above 10 g (the cut-off limit in the El Médano meteorite collection).

We computed a mean saturation activity of 22.9±0.2, 22.6±0.2 and 21.1±0.2 dpm.kg$^{-1}$ for H, L and LL chondrites respectively, values that are consistent to those measured in the literature on meteorite falls (e.g., Nishiizumi et al. 1989, Graf et al. 2001, Dalcher et al. 2013). The hypothesized absence of high shielding is consistent with the small sizes of the studied meteorites (diameter < 10 cm). We derived the uncertainties on terrestrial ages by propagation, which leads to typical range from 60 to 90 ka (Table 1). Finally, we checked that $^{36}$Cl terrestrial production was insignificant to introduce a bias in the results by computing the in situ production at a typical El Médano location (24°40' S, 70°20' W, 2150 m altitude) following the model presented in Schimmelpfennig et al. 2009. This terrestrial $^{36}$Cl production would lead to a saturation activity <10$^{-3}$ dpm.kg$^{-1}$ that is negligible with respect to the $^{36}$Cl activities measured in this work (average 8.30 dpm.kg$^{-1}$, median 6.93 dpm.kg$^{-1}$, minimum 0.060 dpm.kg$^{-1}$).

**ACKNOWLEDGMENTS**

We are indebted to the many colleagues who contributed to the systematic meteorite search conducted in the El Médano area: Leonardo Baeza, Lydie Bonal, Cécile Cournède, Bertrand Devouard, Roger Fu, Albert Jambon, Nejia Lardhidi Ouazaa, Hamed Pourkhorsandi, Neesha Regmi-Schnepf, Minoru Uehara, Maximilien Verdier. We acknowledge funding from the Programme National de Planétologie (PNP, INSU/CNES) and the Agence Nationale de la Recherche (project FRIPON, number ANR13-BS05-0009). This work was also supported by public funds received in the framework of GEOSUD, a project (ANR-10-EQPX-20) of the program "Investissements d'Avenir" managed by the French National Research Agency.MV was supported by the chilean scientific agency CONICYT (FONDECYT projects n°3140562 and

n°11171090). VD thanks Sabrina Cauchies for technical support, and the FRS-FNRS and the ERC StG "ISoSyC" for funding.

**REFERENCES CITED**

Bartholomew, R.B., Boyd, A.W., Brown, F., Hawkings, R.C., Lounsbury, M., and Meritt, W.F., 1955, The half-life of $Cl^{36}$: Canadian Journal of Physics, v. 33, p. 43-48, doi :10.1139/p55-007.

Benoit, P.H., and Sears, D.W.G., 1996, Rapid changes in the nature of the H chondrites falling to Earth: Meteoritics and Planetary Science, v. 31, p. 81-86, doi : 10.1111/j.1945-5100.1996.tb02057.x.

Bland P. A., Sexton A. S., Jull, A. J. T., Bevan, A. WR., Berry, F.J., Thornley, D.M., Astin, T.R., Britt, D.T., and Pillinger, C.T., 1998, Climate and rocks weathering: A study of terrestrial age dated ordinary chondritic meteorites from desert regions: Geochimica et Cosmochimica Acta v. 62, p. 3169–3184, doi: 10.1016/S0016-7037(98)00199-9.

Bland, P.A., Berry, F.J., Smith, T.B., Skinner, S.J., and Pillinger, C.T., 1996, The flux of meteorites to the Earth and weathering in hot desert chondrite finds : Geochimica and Cosmochimica Acta, vol. 60, 11, p. 2053-2059, doi : 10.1016/0016-7037(96)00125-1.

Bottke, W.F., Rubincam, D.P., and Burns, J.A., 2000, Dynamical Evolution of Main Belt Meteoroids: Numerical Simulations Incorporating Planetary Perturbations and Yarkovsky Thermal Forces: Icarus, v. 145, no. 2, p. 301-331, doi : 10.1006/icar.2000.6361.

Bottke, W.F., Durda, D.D., Nesvorny, D., Jedicke, R., Morbidelli, A., Vokrouhlický, D., and Levison, H.F., 2005, Linking the collisional history of the main belt to its dynamical excitation and depletion: Icarus, vol. 179, no. 1, p. 63-94, doi : 10.1016/j.icarus.2005.05.017.

Clarke, J., 2005, Antiquity of aridity in the Chilean Atacama Desert: Geomorphology, v. 73, p. 101-114, doi : 10.1016/j.geomorph.2005.06.008.

Dalcher, N., Caffee, M.W., Nishiizumi, K., Welten, K.C., Vogel, N., Wieler, R., and Leya, I, 2013, Calibration of cosmogenic noble gas production in ordinary chondrites based on 36Cl-36Ar ages. Part 1: Refined produced rates for cosmogenic 21 Ne and 38 Ar: Meteoritics and Planetary Science, v. 48, no. 10, pp. 1841-1862, doi : 10.1111/maps.12203.

Dunai, T.J., Gonzalez-Lopez, G.A, and Juez-Larré, J., 2005, Oligocene-Miocene age of aridity in the Atacama Desert reaveled by exposure dating of erosion-sensitive landforms: Geology, v. 33, no. 4, p. 321-324, doi: 10.1130/G21184.1.

Gattacceca, J., Valenzuela, M., Uehara, M. et al. 2011, The densest meteorite collection area in hot deserts: the San Juan meteorite field (Atacama Desert Chile): Meteoritics and Planetary Science, v.46, no. 9, pp. 1276-1287, doi: 10.1111/j.1945-5100.2011.01229.x.

Genge, M.J., 2008, Koronis asteroid dust within Antarctic ice: Geology , v.36, no. 9, pp. 687-690, doi: 10.1130/G24493A.1

Gladman, B.J., Migliorini, F., Morbidelli, A., Zappalà, V., Michel, P., Cellino, A., Froeschlé, C., Levison, H.F., Bailey, M., and Duncan,M. 1997, Dynamical lifetimes of objects injected into asteroid belt resonances: Science, v. 227, p. 197-201, doi : 10.1126/science.277.5323.197.

Graf, T. and Marti, K., 1995, Collisional history of H chondrites: Journal of Geophysical Research, v. 100, no E10, p. 21,247-21,263, doi : 10.1029/95JE01903.

Graf, T., Caffee, M.W., Martic, K., Nishiizumi, K., and Ponganis, K.V., 2001, Dating Collisional Events: 36Cl-36Ar Exposure ages of H-Chondritic MetalsIcarus, v. 150, no. 1, p. 181-188.

Halliday, I., Blackwell, A.T. and Griffin, A.A., 1989, The flux of meteorites on the Earth's surface: Meteoritics, v. 24, p. 173-178, doi : 10.1111/j.1945-5100.1989.tb00959.x.

Heck, P.R., Schmitz, B., Bottke, W.F., Rout, S.S., Kita, N.T., Cronholm, A., Defouilloy, C., Dronov, A., and Terfelt, F., 2017, Rare meteorite common in the Ordovician period: Nature Astronomy, v. 1, id. 0035, doi : 10.1038/s41550-016-0035

Huss, G.R., 1990, Meteorite infall as a function of mass: Implications for the accumulation of meteorites on Antarctic ice: Meteoritics, v. 25, p. 41-56, doi : 10.1111/j.1945-5100.1990.tb00969.x

Hutzler, A., Gattacceca, J., Rochette, P. et al., 2016, Description of a very dense meteorite collection area in western Atacama: Insight into the long-term composition of the meteorite flux to Earth: Meteoritics and Planetary Science, v. 51, no. 3, pp. 468-482, doi : 10.1111/maps.12607  9.tb00959.x.

Jull, A.J.T.: Terrestrial Ages of Meteorites, in Meteorites and the Early Solar System II, D. S. Lauretta and H. Y. McSween Jr. (eds.), University of Arizona Press, Tucson, 943 pp., p.889-905.

Koblitz, J., 2005, MetBase - Meteorite Data Retrieval Software, version 7.1.

Leya, I., and Masarik, J. 2009,Cosmogenic nuclides in stony meteorites revisited: Meteoritics and Planetary Science, v. 44, no. 7, p. 1061-1086, doi : 10.1111/j.1945-5100.2009.tb00788.x.

Marti, K., and Graf, T., 1992, Cosmic-ray exposure history of ordinary chondrites: Annu. Rev. Earth Planet. Science, no. 20, p. 221-243, doi: 10.1146/annurev.ea.20.050192.001253.

Mazrouei, S., Ghent, R.R., Bottke, W.F., Parker, A.H., and Gernon, T.M.., 2019, Earth and Moon impact flux increased at the end of Paleozoic: Science, v. 363, issue 6424, pp. 253-257, doi: 10.1126/science.aar4058

Merchel, S., and Herpers, U., 1999, An update on radiochemical separation techniques for the determination of long-lived radionuclides via accelerator mass spectrometry: Radiochimica Acta, no. 84, p. 215-220, doi: 10.1524/ract.1999.84.4.215.

Munayco, P., Munayco, J., De Avillez, R.R., Valenzuela, M., Rochette, P., Gattacceca, J., and Scorzelli, R.B., 2013, Weathering of ordinary chondrites from the Atacama Desert, Chile, by Mössbauer spectroscopy and synchrotron radiation X-ray diffraction: Meteoritics And Planetary Science no. 48, p. 457-473, doi: 10.1111/maps.12067.

Nishiizumi, K., Elmore, D., and Kubik, P.W., 1989, Update on terrestrial ages of Antarctic meteorites: Earth and Planetary Science Letters, v. 93, no. 3-4, pp. 299-313, doi : 10.1016/0012-821X(89)90029-0.

Schimmelpfennig, I., Benedetti, L., Finkel, R., Pik, R., Blard, P.H., Bourlès, D., Burnard, P, and Williams, A., 2009, Source of in-situ 36Cl in basaltic rocks. Implications for calibration of production rates: Quaternary geochronology, v. 4, p. 441-461, doi : 10.1016/j.quageo.2009.06.003.

Schmitz, B., Tassinari, M., and Peucker-Ehrenbrink, B., 2001, A rain of ordinary chondritic meteorites in the early Ordovician: Earth and Planetary Science Letters, v. 194, issue 1-2, p. 1-15, doi : 10.1016/S0012-821X(01)00559-3

Schmitz, B., 2013, Extraterrestrial spinels and the astronomical perspective on Earth's geological record and evolution of life: Chimie der Erde, v. 73, issue 2, pp. 117-145, doi : 10.1016/j.chemer.2013.04.002

Vogt, S., and Harpers, U., 1988, Radiochemical separation techniques for the determination of long-lived radionuclides in meteorites by means of accelerator mass-spectroscopy: Fresenius Z. Anal. Chem., no. 331, pp. 186-188, doi: 10.1007/BF01105164.

Whillans, I.M. and Cassidy, W.A. et al., 1983, Catch a falling star: meteorites and old ice: Science, v. 7, no. 222, pp. 55-57, doi: 10.1126/science.222.4619.55.

Welten, K.C., Nishiizumi, K., Caffee, M.W, Hillegonds, D.J., Johnson, J.A., Jull, A.J.T., Wieler, R., and Folco, L., 2006, Terrestrial ages, pairing, and concentration mechanism of Antarctic chondrites from Frontier Moutain, Northern Victoria Land: Meteoritics and Planetary Science, v. 41, no. 7, pp. 1081-1094, doi: 10.1111/j.1945-5100.2006.tb00506.x.

Wiegert, P. and Brown, P., 2005, The Quadrantid meteoroid complex: Icarus, v. 179, Issue 1, pp. 139-157, doi: 10.1116/j.icarus.2005.05.019

Zolensky, M., 1998, The flux of meteorites to Antarctica, Geological Society - London Special Publications, 140, pp. 93-104, doi: 10.1144/GSL.SP.1998.140.01.09

Zolensky, M., Bland, P., Brown, P., and Halliday, I., 2006: Flux of Extraterrestrial Materials, in Meteorites and the Early Solar System II, D. S. Lauretta and H. Y. McSween Jr. (eds.), University of Arizona Press, Tucson, 943 pp., p.869-888.